\documentclass[review,11pt]{elsarticle}
\usepackage{graphicx}
\usepackage{lscape}
\usepackage{upgreek}
\usepackage{caption} 
\usepackage{hyperref}
\usepackage{todonotes}
\usepackage{amssymb} 
\usepackage{lineno}
\usepackage{float}
\usepackage[normalem]{ulem}
\usepackage{amsmath}
\usepackage{subcaption}
\usepackage{tcolorbox}

\usepackage{longtable}
\usepackage{multirow}
\usepackage{soul}
\usepackage{algorithmic}

\newcommand{\quickwordcount}{
  \immediate\write18{texcount -1 -sum -merge \jobname.tex > \jobname-words.sum }
  \input{\jobname-words.sum} words
}
\usepackage{tikz}
\definecolor{darkpastelpurple}{rgb}{0.59, 0.44, 0.84}
\definecolor{darkmagenta}{rgb}{0.55, 0.0, 0.55}

\tikzstyle{mybox} = [draw=black, very thick, rectangle, rounded corners, inner ysep=5pt, inner xsep=5pt]
\usepackage{color}

\newcommand\rs[1]{\textcolor{black}{#1}}
\newcommand\rff[1]{\textcolor{black}{#1}}
\newcommand\rss[1]{\textcolor{black}{#1}}

\usepackage[title]{appendix}

\newcommand{\rb}[1]{
	
	\begin{tcolorbox}[colback=gray!05,
		colframe=black,
		width=\columnwidth,
		arc=3mm, auto outer arc,
		boxrule=0.5pt,
		]
		#1
	\end{tcolorbox}
}

\newcounter{Finding}
\stepcounter{Finding}

\newcommand{\roundedbox}[1]{
	\rb{
		\noindent
 		\textit{#1}
	}
	\stepcounter{Finding}
}

\RequirePackage[normalem]{ulem} 
\RequirePackage{color}\definecolor{RED}{rgb}{1,0,0}\definecolor{BLUE}{rgb}{0,0,1} 
\providecommand{\DIFaddbegin}{} 
\providecommand{\DIFaddend}{} 
\providecommand{\DIFdelbegin}{} 
\providecommand{\DIFdelend}{} 
\providecommand{\DIFaddbeginFL}{} 
\providecommand{\DIFaddendFL}{} 
\providecommand{\DIFdelbeginFL}{} 
\providecommand{\DIFdelendFL}{} 
\newcommand{\DIFscaledelfig}{0.5}
\RequirePackage{settobox} 
\RequirePackage{letltxmacro} 
\newsavebox{\DIFdelgraphicsbox} 
\newlength{\DIFdelgraphicswidth} 
\newlength{\DIFdelgraphicsheight} 
\LetLtxMacro{\DIFOincludegraphics}{\includegraphics} 
\newcommand{\DIFaddincludegraphics}[2][]{{\color{blue}\fbox{\DIFOincludegraphics[#1]{#2}}}} 
\newcommand{\DIFdelincludegraphics}[2][]{
\sbox{\DIFdelgraphicsbox}{\DIFOincludegraphics[#1]{#2}}
\settoboxwidth{\DIFdelgraphicswidth}{\DIFdelgraphicsbox} 
\settoboxtotalheight{\DIFdelgraphicsheight}{\DIFdelgraphicsbox} 
\scalebox{\DIFscaledelfig}{
\parbox[b]{\DIFdelgraphicswidth}{\usebox{\DIFdelgraphicsbox}\\[-\baselineskip] \rule{\DIFdelgraphicswidth}{0em}}\llap{\resizebox{\DIFdelgraphicswidth}{\DIFdelgraphicsheight}{
\setlength{\unitlength}{\DIFdelgraphicswidth}
\begin{picture}(1,1)
\thicklines\linethickness{2pt} 
{\color[rgb]{1,0,0}\put(0,0){\framebox(1,1){}}}
{\color[rgb]{1,0,0}\put(0,0){\line( 1,1){1}}}
{\color[rgb]{1,0,0}\put(0,1){\line(1,-1){1}}}
\end{picture}
}\hspace*{3pt}}} 
} 
\LetLtxMacro{\DIFOaddbegin}{\DIFaddbegin} 
\LetLtxMacro{\DIFOaddend}{\DIFaddend} 
\LetLtxMacro{\DIFOdelbegin}{\DIFdelbegin} 
\LetLtxMacro{\DIFOdelend}{\DIFdelend} 
\DeclareRobustCommand{\DIFaddbegin}{\DIFOaddbegin \let\includegraphics\DIFaddincludegraphics} 
\DeclareRobustCommand{\DIFaddend}{\DIFOaddend \let\includegraphics\DIFOincludegraphics} 
\DeclareRobustCommand{\DIFdelbegin}{\DIFOdelbegin \let\includegraphics\DIFdelincludegraphics} 
\DeclareRobustCommand{\DIFdelend}{\DIFOaddend \let\includegraphics\DIFOincludegraphics} 
\LetLtxMacro{\DIFOaddbeginFL}{\DIFaddbeginFL} 
\LetLtxMacro{\DIFOaddendFL}{\DIFaddendFL} 
\LetLtxMacro{\DIFOdelbeginFL}{\DIFdelbeginFL} 
\LetLtxMacro{\DIFOdelendFL}{\DIFdelendFL} 
\DeclareRobustCommand{\DIFaddbeginFL}{\DIFOaddbeginFL \let\includegraphics\DIFaddincludegraphics} 
\DeclareRobustCommand{\DIFaddendFL}{\DIFOaddendFL \let\includegraphics\DIFOincludegraphics} 
\DeclareRobustCommand{\DIFdelbeginFL}{\DIFOdelbeginFL \let\includegraphics\DIFdelincludegraphics} 
\DeclareRobustCommand{\DIFdelendFL}{\DIFOaddendFL \let\includegraphics\DIFOincludegraphics} 

\let\today\relax
\makeatletter
\def\ps@pprintTitle{%
    \let\@oddhead\@empty
    \let\@evenhead\@empty
    \def\@oddfoot{\footnotesize\itshape
         {
         \hspace*{-3cm}
         \includegraphics[width=1.46\linewidth,trim={{1.7cm} {2cm} {1.2cm} {24cm}},clip]{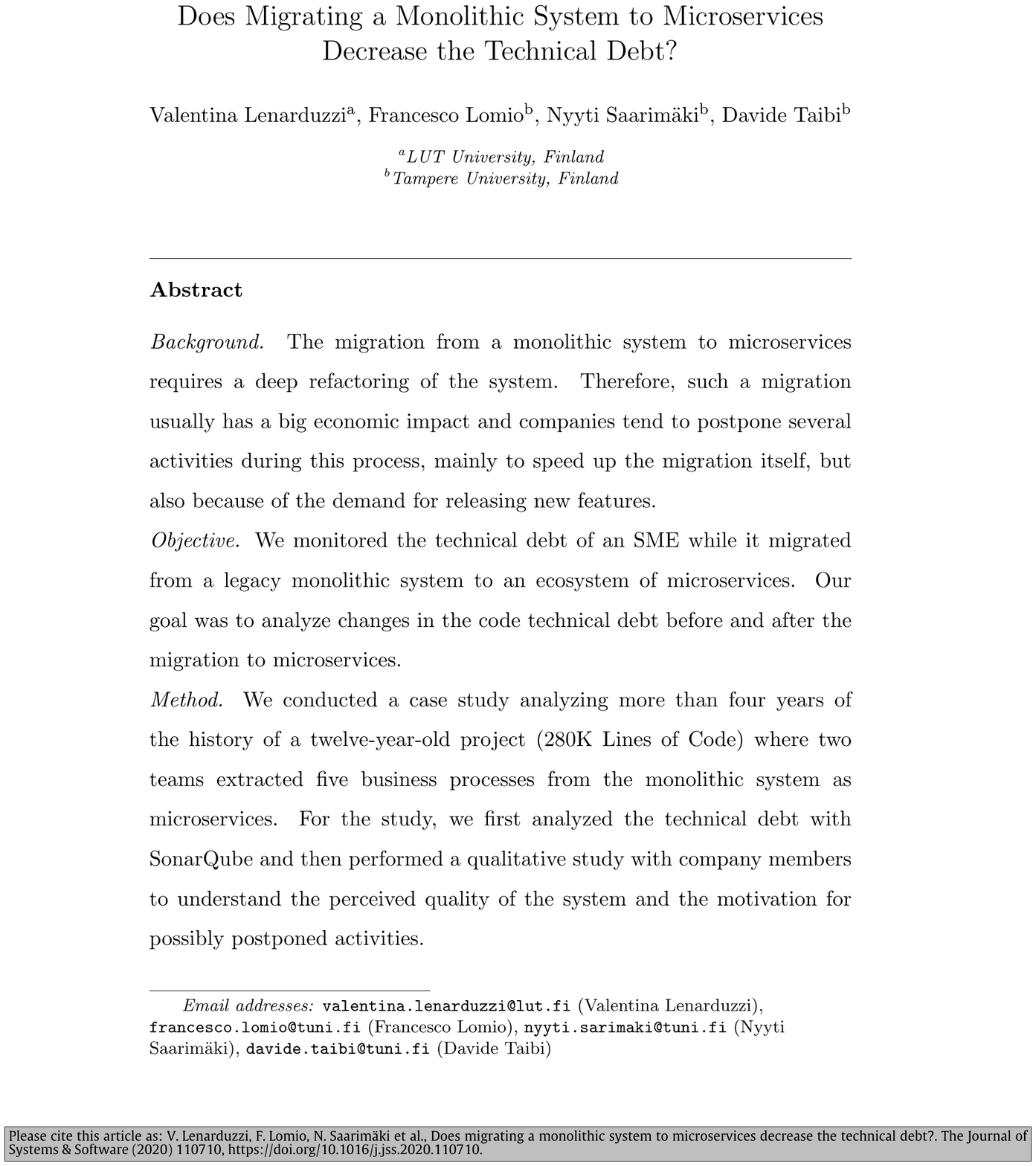}} \hfill\today}%
    \let\@evenfoot\@oddfoot
    }
\makeatother

\begin{document}
\begin{frontmatter}

\title{Does Migrating a Monolithic System to Microservices Decrease the Technical Debt? }

\author [LUT] {Valentina Lenarduzzi}
\ead{valentina.lenarduzzi@lut.fi}

\author [TUNI]{Francesco Lomio}
\ead{francesco.lomio@tuni.fi}

\author [TUNI]{Nyyti Saarim{\"a}ki}
\ead{nyyti.sarimaki@tuni.fi}

\author [TUNI]{Davide Taibi}
\ead{davide.taibi@tuni.fi}

\address [LUT] {LUT University, Finland}
\address [TUNI] {Tampere University, Finland}

\begin{abstract}
\textit{Background.} The migration from a monolithic system to microservices requires a deep refactoring of the system. Therefore, such a migration usually has a big economic impact and companies tend to postpone several activities during this process, mainly to speed up the migration itself, but also because of the demand for releasing new features.  
\\
\textit{Objective.} We monitored the technical debt of an SME while it migrated from a legacy monolithic system to an ecosystem of microservices. Our goal was to analyze changes in the code technical debt before and after the migration to microservices. \\
\textit{Method.} We conducted a case study analyzing more than four years of the history of a twelve-year-old project (280K Lines of Code) where two teams extracted five business processes from the monolithic system as microservices. For the study, we first analyzed the technical debt with SonarQube and then performed a qualitative study with company members to understand the perceived quality of the system and the motivation for possibly postponed activities.  \\
\textit{Results.} The migration to microservices helped to reduce the technical debt in the long run. Despite an initial spike in the technical debt due to the development of the new microservice, after a relatively short period of time the technical debt tended to grow slower than in the monolithic system. \\
\end{abstract}

\begin{keyword}
Technical Debt\sep
Architectural Debt\sep
Code Quality\sep
Microservices\sep
Refactoring
\end{keyword}

\end{frontmatter}


\section{Introduction}
\label{Intro}

Migration to microservices has become very popular over the last years. Companies migrate for different reasons, for example because they expect to improve the quality of their system or to facilitate software maintenance~\cite{Taibi2017}. 

Companies commonly adopt an initial migration strategy to extract components from their monolithic system as microservices, making use of simplified microservices patterns~\cite{Taibi2017}\cite{TaibiCLOSER}. As an example, it is common for companies to initially connect the microservices directly to the legacy monolithic system and not to adopt message buses. However, when the system starts to grow in complexity, they usually start re-architecting their system, considering different architectural patterns ~\cite{Taibi2017}\cite{TaibiCLOSER}\cite{Knoche2018}. The migration from a monolithic system to microservices is commonly performed on systems that are being actively developed. Therefore, in several cases, the development of new features is prioritized over refactoring of the code, which generates technical debt (TD) every time an activity is postponed~\cite{Cunningham1992}\cite{Li2015} and increases the cost of software maintenance. 

Companies migrate to microservices to facilitate maintenance~\cite{Taibi2017}. However, recent surveys have confirmed that maintenance costs increase after migration~\cite{Taibi2017}\cite{SOLDANI2018215}. 
Therefore, the goal of this paper is to understand whether the overall code TD of a system increases or decreases after migration to microservices, and whether the type of TD varies after the migration.

In this work, we report a case study where we monitored the code TD of an SME (small and medium-sized enterprise) that migrated their legacy monolithic system to an ecosystem of microservices. 
 The company was interested in evaluating their TD with SonarQube\footnote{SonarQube: www.sonarqube.org}, which is one of the most commonly used tools for analyzing code TD. They asked us to monitor the evolution of TD in a project they are developing, during the migration to microservices.
The analysis focused on the three types of TD proposed by SonarQube: reliability remediation cost (time to remove all the issues that can generate faults), maintainability remediation cost (time to remove all the issues that increase the maintenance effort), and security vulnerability remediation cost (time to remove all the security issues). Moreover, we implemented a qualitative study by conducting a focus group with two development teams, the software architect, and the product manager. The goal was to deeply understand the causes of the changes in the distribution of the types of TD issues and the motivations for any postponed activities.

To the best of our knowledge, only a limited number of studies have investigated the impact of postponed activities on TD, especially in the context of microservices~\cite{DeToledo2019}. This work will help companies to understand how TD grows and changes over time while at the same time opening up new avenues for future research on the analysis of TD interest.

This paper is structured as follows: Section~\ref{Background} briefly introduces the background and related work on microservices and TD, \rs{while Section~\ref{RW} reports the existing work in this topic}. In  Section~\ref{CS}, we present the case study design, defining the research questions and describing the study context with the data collection and the data analysis protocol. In Section~\ref{Results}, we show the results we obtained, followed by a discussion of them in Section~\ref{Discussion}. In Section~\ref{Threats}, we identify the threats to the validity of our study, and in Section~\ref{Conclusion}, we draw conclusions and provide an outlook on possible future work. 
\section{Background}
\label{Background}

In this Section, we will first describe the background on microservices and technical debt (TD). Moreover, we will describe SonarQube and the method adopted to calculate TD.

\subsection{Microservices}
Microservice architecture has become more and more popular over the last years. 
Microservices are small, autonomous, and independently deployed services, with a single and clearly defined purpose \cite{Newman}. 

The independent deployment provides a lot of advantages. They can be developed in different programming languages, they can scale independent of other services, and they can be deployed on the hardware that best suits their needs. Moreover, because of their size, they are easier to maintain and more fault-tolerant since the failure of one service will not break the whole system, which is possible in a monolithic system.
In addition, as microservices are cloud-native applications, they support the IDEAL properties: Isolation of state, Distribution, Elasticity, Automated management, and Loose Coupling \cite{Newman}. Moreover, microservices propose vertical decomposition of applications into a subset of business-driven services. Each service can be developed, deployed, and tested independently by a different development team using different technology stacks. The development responsibility of a microservice belongs to a single team, which is in charge of the whole development process, including deploying, operating, and upgrading the service when needed.

\subsection{Technical Debt}
The concept of TD was introduced for the first time in 1992 by Cunningham as \textit{''The debt incurred through the speeding up of software project development which results in a number of deficiencies ending up in high maintenance overheads''}~\cite{Cunningham1992}.  McConnell~\cite{McConnell2013} improved the definition of TD to \textit{''A design or construction approach that's expedient in the short term but that creates a technical context in which the same work will cost more to do later than it would cost to do now (including increased cost over time)''}. 
In 2016, Avgeriou et al.~\cite{Avgeriou2016} defined TD as \textit{''A collection of design or implementation constructs that are expedient in the short term, but set up a technical context that can make future changes more costly or impossible. TD presents an actual or contingent liability whose impact is limited to internal system qualities, primarily maintainability and evolvability''}.

Different approaches and strategies have been suggested for evaluating TD. Nugroho et al.~\cite{Nugroho2011} proposed an approach for quantifying debt based on the effort required to fix TD issues, using data collected from 44 systems as their basis.
Seaman et al.~\cite{Seaman2011} proposed a TD management framework that formalizes the relationship between cost and benefit in order to improve software quality and support the decision-making process during maintenance activities.
Zazworka et al.~\cite{Zazworka2013} investigated automated identification of TD. They asked developers to identify TD items during the development process and compared the manual identification with the results from an automatic detection.
Zazworka et al.~\cite{Zazworka2014} examined source code analysis techniques and tools to identify code debt in software systems, focusing on TD interest and TD impact on increasing defect- and change-proneness.
They applied four TD identification techniques (code smells, automatic static analysis issues, grime buildup, and modularity violations) on 13 versions of the Apache Hadoop open-source software project. They collected different metrics, such as code smells and code violations. The results showed a positive correlation between some metrics and defect- and change-proneness, such as Dispersed Coupling and modularity violations. 
Guo et al.~\cite{Guo2016} investigated the TD cost of applying a new approach to an ongoing software project. They found higher start-up costs, which decreased over time. 


\subsubsection{Technical Debt Measurement}

Different commercial and open-source tools can be used to measure TD, including CAST\footnote{CAST Software
https://www.castsoftware.com/ Last Access: August 2019}, Coverity Scan\footnote{Coverity Scan. https://scan.coverity.com. Last Access: August 2019}, SQUORE\footnote{SQUORE. https://www.squoring.com/en/produits/squore-software-analytics/ \\Last Access: August 2019}, and Designite\footnote{Designite. http://designite-tools.com Last Access: August 2019}.

In this work, as required by our case company, we adopted SonarQube, as it is one of the most commonly used TD measurement tools, adopted by more than 120K users\footnote{SonarQube. http://www.sonarqube.org Last Access: August 2019 }. Moreover, SonarQube is also open-source, while the other well-known competitors have a commercial license. 

SonarQube calculates several metrics such as number of lines of code and code complexity, and verifies the code's compliance against a specific set of ''coding rules''. If the analyzed source code violates a coding rule or if a metric is outside a predefined threshold (also called "quality gate"), SonarQube generates an "issue". 
 Issues are problems that generate TD and therefore should be solved.

SonarQube has separate rule sets for the most common development languages such as Java, Python, C++, and JavaScript. For example, SonarQube version 7.5 includes more than 500 rules for Java. 

Rules are classified as being related to reliability, maintainability, or security of the code. Reliability rules, also named "bugs", create TD issues that ''represent something wrong in the code'' and that will soon be reflected in a bug. Security rules, also called "vulnerabilities" or "security hotspots", represent issues that can be exploited by a hacker or that are otherwise security-sensitive. Maintainability rules or ''code smells'' are considered ''maintainability-related issues'' in the code that decrease code readability and modifiability. It is important to note that the term ''code smells'' adopted in SonarQube does not refer to the commonly known code smells defined by Fowler et al.~\cite{Fowler1999}, but to a different set of rules. SonarQube claims that zero false-positive issues are expected from the reliability and maintainability rules, while security issues may contain some false-positives\footnote{SonarQube Rules: https://docs.sonarqube.org/display/SONAR/Rules \\ Last Access: June 2019}. The complete list of rules is available online\footnote{https://rules.sonarsource.com/java Last Access: August 2019}. 

SonarQube calculates TD using the SQALE method~\cite{mordal2009squale}. It is an ISO 9126 compliant method 
developed by DNV ITGS France~\cite{letouzey2012managing}. The method is based on five categories~\cite{curtis2012estimating}:
\begin{itemize}
    \item Robustness: Application stability and ability to recover from failures.
	\item Performance efficiency: Application responsiveness and usage of resources.
	\item Security: System’s ability to protect confidential information and prevent unauthorized intrusions.
	\item Transferability: Software understandability and especially the ''ease  with which a new team can understand the software and become productive''.
	\item	Changeability:  Measures software adaptability and modifiability.
\end{itemize}{}

SonarQube calculates TD as:
\begin{equation}
\begin{split}
\centering
 TD_{(person-days)}= & cost\;to\;fix\;issues + cost\;to\;fix\;duplications  + cost\;to\;comment\;\\* & public\;API  +
cost\; to\;fix\;uncovered\;complexity + cost\;to\;bring \\* & \;complexity\;below\;threshold   
\end{split}
\end{equation}

SonarQube also identifies TD density as: 
\begin{equation}
TD Density = \frac{TD_{(person-days)}}{KLOC} 
\end{equation}

The previous formula is then instantiated three times to calculate: 
\begin{itemize}
    \item \textit{Technical Debt}, considering maintainability-related rules (tagged as ''code smells'') that are supposed to increase the change-proneness of the infected code. 
    \\Technical Debt was also called ''SQUALE index'' until SonarQube version 7.7. Starting from SonarQube version 7.8, it has been called ''Mantainability remediation effort''. In this work, we refer to this type of TD as TD\_M.
    \item \textit{Reliability Remediation Effort}, considering reliability-related rules (rules tagged as ''bug'') that are supposed to increase the bug-proneness of the infected code. In this work, we refer to this type of TD as TD\_R.
    \item \textit{Security Remediation Effort}, considering rules related to security vulnerabilities (tagged as ''vulnerability''). In this work, we refer to this type of TD as TD\_S. 
\end{itemize}

\section{Related Work}
\label{RW}
In this Section, we report on the most relevant related work on migration to microservices and TD. 

Companies migrate to microservices in order to ease their software development by improving maintainability and decreasing delivery time~\cite{Taibi2017}\cite{SOLDANI2018215}.
However, migration to microservices is not an easy task. Companies commonly start this migration without having any experience with microservices, and only in few cases do they hire a consultant to support them during the migration~\cite{Taibi2017}\cite{SOLDANI2018215}. 
\rff{\subsection{Migration from Monolithic to Microservice}}
\rff{Several approaches have been proposed for migrating from a monolithic system to microservices. }

Fritzsch et al.~\cite{Fritzsch2019} analyzed works from the literature and classified the reported refactoring approaches used to migrate from monolithic systems to microservices. They highlight that not all of the existing refactoring approaches are practically applicable, nor do they provide adequate tool support and metrics to verify the results of the migration.

In our previous work~\cite{Taibi2017}, we classified the different migration processes adopted by companies, highlighting the complex migration steps and the need of support from an experienced software architect, at least for identifying the architectural guidelines and helping to initiate the migration. 

Lu et al.~\cite{Lu2019} identified other migration strategies, distinguishing between the big bang migration, where companies replace their legacy system in one swoop, and incremental migration, where companies do the replacement step by step. 

\rff{Abbott and Fischer~\cite{Abbott2015} proposed a decomposition approach based on the ''scalability cube'', which splits an application into smaller components to achieve higher scalability. Richardson~\cite{Richardson2017} also mentioned this approach in his four decomposition strategies:} 
\rff{\begin{itemize}
    \item ''Decompose by business capability and define services corresponding to business capabilities'';
    \item ''Decompose by domain-driven design sub-domain'';
    \item ''Decompose by verb or use ‘cases’ and define services that are responsible for particular actions'';
    \item ''Decompose by nouns or resources by defining a service that is responsible for all operations on entities/resources of a given type''.
\end{itemize}}
 
\rff{Kecskemeti et al.~\cite{Kecskemeti2016} proposed a decomposition approach based on container optimization. The goal is to increase the elasticity of large-scale applications and the possibility to obtain more flexible compositions with other services. }


\rff{Vresk et al.~\cite{VreskC16} recommend combining verb-based and noun-based decomposition approaches. The proposed approach hides the complexity stemming from the variation of end-device properties thanks to the application of a uniform approach for modeling both physical and logical IoT devices and services. Moreover, it can foster interoperability and extensibility using diverse communication protocols into proxy microservice components.} 

\rff{Gysel et al.~\cite{Gysel16} proposed a clustering algorithm approach based on 16 coupling criteria derived from literature analysis and industry experience. This approach is an extensible tool framework for service decomposition as a combination of a criteria-driven methods. It integrates graph clustering algorithms and features priority scoring and nine types of analysis and design specifications. Moreover, this approach introduces the concept of coupling criteria cards using 16 different instances grouped into four categories: cohesiveness, compatibility, constraints, and communications. The approach was evaluated by integrating two existing graph clustering algorithms, combining actions research and case study investigations, and load tests. The  results showed potential benefits to the practitioners, also confirmed by  user feedback. }

\rff{Chen et al.~\cite{Chen2017} proposed a data-driven microservice-oriented decomposition approach based on data flow diagrams from business logic. Their approach could deliver more rational, objective, and easy-to-understand results thanks to objective operations and data extracted from real-world business logic. Similarly, we adopt process mining to analyze the business processes of a monolithic system.}

\rff{Alwis et al.~\cite{Alwis2018} proposed a heuristic to slice a monolithic system into microservices based on object sub-types (i.e., the lowest granularity of software based on structural properties) and functional splitting based on common execution fragments across software (i.e., the lowest granularity of software based on behavioral properties).}

\rff{Stojanovic et al.~\cite{Stojanovic2020}, recently proposed an approach to identify microservices in monolithic systems using structural system analysis.}

\rff{Other studies~\cite{Cojocaru2019, Rud2006, Zdun2017, Engel2018} proposed different approaches to validate the migration from monolithic to microservices. }

\rff{\subsection{Microservices and Technical Debt}}


Several works have investigated TD in microservices.

\rff{Chhonker and de Lemos~\cite{Chhonker2019} proposed a work plan to investigate the impact  of TD on microservices, adopting   self-adaption as a solution to reduce the TD of microservice-based software systems.}

As for possible issues that can generate TD in microservice-based systems, in our previous works~\cite{taibiIEEE2018}\cite{TaibiAntipatternBook2019}, we identified a set of microservice-specific anti-patterns  and ''bad smells'' that can cause TD in microservice-based systems. Recently, Bogner et al.~\cite{BognerAntiPattern2019} extended our work, creating a public catalog of anti-patterns. 

Bogner et al.~\cite{BognerTD2018} ran an industrial survey investigating the approaches adopted by industry to prevent the accumulation of TD, reporting that companies do have problems to prevent TD due to architectural erosion, mainly because of the lack of automated quality control at the architectural level. Moreover, in a subsequent study, Bogner et al.~\cite{BognerICSME2019} performed another industrial survey, interviewing 17 practitioners to  explore the evolvability assurance processes applied and the usage of tools, metrics, and patterns in microservices-based systems. They reported that architectural issues, and especially postponed architectural decisions, are the most harmful type of TD. Moreover, they also reported that their participants did not apply any architectural or service-oriented tools or metrics. As a result, they recommend applying static analysis tools and architectural analysis tools to keep track of the software quality and especially of the architectural issues in the systems. 

In their study, de Toledo et al.~\cite{DeToledo2019} performed an exploratory case study on a large industry/company while it was refactoring an existing microservice-based system, removing issues in the communication layer. They focused their investigation on the TD related to the communication layer of the system, indicating the large number of point-to-point connections between services as the major issue and noting the presence of business logic in the communication layer, which increased the dependency between services. Similarly to this study, we performed a case study on a single company, but we focused on the migration process itself, analyzing the system in all its aspects before and after the migration. 

M\'arquez and Astudillo~\cite{Marquez2018} proposed an approach for identifying architectural TD in monolithic systems before the migration in order to satisfy the requirements of the new microservices-based system. Their proposal includes a set of tactics and patterns for better contextualizing the different types of architectural TD. 


\rff{Rademacher et al.~\cite{Rademacher2019}, introduce an approach to make technology decisions in microservice architectures explicit and enable reasoning about them, also enabling to be aware of the  TD accrued by the different decisions.}

\rff{Brogi et al.~\cite{Brogi2020} propose a methodology to reduce TD in microservices by systematically identifying the architectural smells that violate the main design principles of microservices, and to select suitable architectural refactorings to resolve them. }

Differently than in the previous works, \rss{in this work we are comparing TD and its trend in a project before and after migration to microservices, considering both the code TD detected by SonarQube and the TD perceived by the developers.}

\section{Case Study Design}
 \label{CS}
We designed our empirical study as a case study based on the guidelines defined by Runeson and H\"{o}st~\cite{Runeson2009}.
In this Section, we will describe the case study design, including the goal and the research questions, the study context together with the case and subject selection, the data collection, and the data analysis procedure.

\subsection{Goal and Research Questions}
\label{RQ}
In this case study, we compared technical debt (TD) and its trend in a project before and after migration to microservices. The study was performed based on the following research questions (\textbf{RQ$_s$}): \\

\begin{itemize}
    \item [\textbf{RQ$_1$:}] Is the TD of a monolithic system growing with the same trend as a microservices-based system?

    \begin{itemize}
        \item [\rss{RQ$_{1.1}$:}] \rss{Is the \underline{TD} of a monolithic system, \underline{measured with the SonarQube} \textit{"sonar way"} model, growing with the same trend as a microservices-based system?}

        \item [\rss{RQ$_{1.2}$:}] \rss{Is the TD \underline{perceived by the developers} growing with the same trend in the monolithic and in the microservice-based systems?} 
    \end{itemize} 
    
    \item [\rss{\textbf{RQ$_2$:}}] \rss{How the different types of TD change after the migration to microservices?}

    \begin{itemize}
        \item [\rss{RQ$_{2.1}$:}] How the \underline{different types of TD issues,  measured by SonarQube} (bugs, code smells, and security vulnerabilities), change after the migration to microservices?
    
        \item [RQ$_{2\rss{.2}}$:]  \rss{How the \underline{different types of perceived TD} change after the  migration to microservices?}
    \end{itemize}
\end{itemize}

\rss{With \textbf{RQ$_1$} we aim at understanding if the overall TD is growing with the same trend in a monolithic and in a microservice-based system. Since our case company is interested in evaluating TD with SonarQube, we investigate the TD calculated at code level by SonarQube (\textbf{RQ$_{1.1}$}). However, in order to get more insights on the results obtained for the execution of SonarQube, we investigate if the TD perceived by the developers (\textbf{RQ$_{1.2}$}) also changes after the migration to microservices. 
We \textit{hypothesize} that the TD decreases after the migration to microservices, mainly because of the newly developed architectural design and the re-developed code that should be written in a more understandable way.} 

\rss{In \textbf{RQ$_2$} we investigate if different types of TD,  or different related qualities, are also affected by the migration. As an example, we expect the code to be less buggy after the migration to microservices, but also to be more stable and less change-prone. As well as for RQ1, we first investigate changes in the different types of TD items detected by SonarQube (\textbf{RQ$_{2.1}$}) and then we investigate if the developers perceive changes on the same qualities proposed by SonarQube (\textbf{RQ$_{2.2}$}). Therefore, we \textit{hypothesize} that the microservice-based system should be less faulty (less SonarQube issues of type ''Bugs''), less change-prone (SonarQube issues of type ''Code Smells'') and more secure (SonarQube issues of type ''Security Vulnerabilitiyes'').  
}










\subsection{Case and Subject Selection}
\label{Context}
\rss{In this Section, we describe the case company of this study, the system that has been considered for investigating our RQs, and the process for migrating to microservices adopted by the case company}


\textbf{The Case Company}.
The company is an Italian Small and Medium-Sized Enterprise (SME) with 50 developers, developing different business suites for tax accountants, lawyers, and other related businesses. The developed systems are used by more than 10K practitioners in Italy. 

\textbf{The System Under Migration}.
The company is migrating the bookkeeping document management system for Italian tax accountants. The \rff{system allows a tax accountant to } manage the whole bookkeeping process, including management of the digital invoices, send the invoices to the Ministry of Economic Development, manage tax declarations, and fulfill all legal requirements. The system is currently being used by more than 2,000 tax accountants, who need to store more than 20M invoices per year. 

The company needs to frequently update the system, as the annual tax rules usually change every year. The Italian government normally updates the bookkeeping process between December and January, which involves not only changing the tax rate but also modifying the process of storing the invoices. However, tax declarations can be made starting in March/April of each year. Therefore, in the best case, the company has two to four months to adapt their software to the new rules in order to enable tax accountants to work with the updated regulations from March/April. Up to now, the company needed to hire a consultancy company to help them during these three months of fast-paced work. 

\rff{ The system is being developed as a Java J2EE application. The system is based on a server-side monolith, developed with a Model-View-Controller pattern, a web application developed with a set of Java Servlets and jsp pages, and a desktop application based on Java Swing to support document uploading and synchronization. The server-side component is deployed as a single war file into a Tomcat 6 server, running on a virtual machine in Microsoft Azure Cloud. Data is stored on a Microsoft SQL server database.
}

 The project is being developed by three teams: two teams composed of four developers each, and one team of five consultants usually hired from December to May to help with adapting the system to the new tax rules.  
 
 The system has been growing every year and as a result, \rff{after twelve years and more than 280K of lines of code}, the code has gotten harder to understand and especially to modify. The CEO reported that before the migration, changes took a significant amount of time to implement, as they had to modify the code in several places. Moreover, every time new developers joined in the company, they took a long time to understand the system and to be able to implement changes. 

As a result of the aforementioned issues, the company decided to migrate to microservices to facilitate maintenance of the system and to distribute the work to independent groups by separating each business process, eliminating the need for synchronization between teams, and increasing velocity.

\rss{The company migrated five business processes to independent microservices, which had been extracted from the monolithic system.}

\rss{The microservices were also developed in Java, and automatically deployed as Docker containers on a Kubernetes cluster in the Microsoft Azure Cloud.}

\rss{Two microservices were responsible of the upload and the permanent storage of the invoices, one was responsible for managing the tax declarations from the private database, and two were implemented to store and manage the tax declarations in the portal of the Ministry of Economic Development. The reason of implementing both the private and public management of the invoices, is that tax accountants need to store their private copy of the tax declarations, independently from the data stored on the portal of the Ministry of Economic Development.}

\rss{After the development of the first two microservices, the teams adopted RabbitMQ as message bus, for managing communications between the microservices and the monolithic system. Then, after the implementation of the third microservice, they started using RabbitMQ also as an API-Gateway.   
}

\rss{The five microservices are functionally different, but structurally similar. All of them are written in the same language (Java) and their main purpose is to manage data from invoices or tax declarations and store or access them in different formats and from different locations (local database, SQL Server, or the APIs of the Ministry of Economic Development).}


\subsection{Study Execution and Data Collection}
\label{DataCollection}

We monitored the migration process from August 2014 to September 2018. 

The study was performed in two steps. 
\rff{\begin{itemize}
    \item  \textit{Automated TD analysis}: In this step, we  collected data on code TD in system,  before and during the migration, to answer our \textbf{RQ1.1}.
    \item \textit{Focus group } to understand the usefulness of the TD analysis and gather more qualitative insights  (\textbf{RQ2.1)}
\end{itemize}
}

\newpage
\textbf{Automated TD Analysis}.

The TD data was obtained by analyzing the system's commits using SonarQube\footnote{https://www.sonarqube.org} (version 7.0). For the analysis, we used SonarQube's standard quality profile and analyzed each commit two years before the migration and after the start of the migration.

We analyzed the TD provided by SonarQube considering the distribution of the following types of TD issues: 
\begin{itemize}
    \item \textbf{TD\_M}: SonarQube ''Technical Debt'', also called ''Maintainability Remediation Effort'' (issues classified by SonarQube as "Code Smells")
    \item \textbf{TD\_R}: \rs{Reliability} Remediation Effort (issues classified as ''Bugs'')
    \item  \textbf{TD\_S}: Security Remediation Effort (issues classified as ''Security Vulnerabilities'')
\end{itemize}

\textbf{Focus Group}. To confirm the results of the automated TD analysis and gain more qualitative insights on the results, we performed a focus group as the second step. 
The focus group was based on a face-to-face semi-structured interview. The goal was to discuss whether the migration to microservices had been beneficial and whether the participants had experienced any of the expected benefits. 
The focus group was moderated by one of the authors. The whole focus group session was audio-recorded. To be able to do this, we obtained a written consent from all the participants. The relevant parts were transcribed and then coded by the moderator of the focus group, and verified by one of the authors.

During the focus group, we asked four questions: \\

\noindent\textit{\textbf{Q1:} What benefits and issues have you perceived after the migration to microservices?} Please write the issues on the red Post-it notes and the benefits on the yellow ones. 
\\The goal of this question was to understand whether the practitioners perceived some benefits that can be associated to the change of TD.  \\
\noindent\textit{\textbf{Q2:} Is the quality of the code changed after the migration to microservices? } \\
This question aimed at understanding the perception of the overall project quality by the developers before and during the migration. 

\noindent\textit{\textbf{Q3:} Has the TD decreased after the migration to microservices?}  \\
Before asking this question, the moderator introduced the notion of TD. 
The goal of this question was to understand the perceived overall TD from the developers' point of view. 

\noindent\textit{\textbf{Q4:} Have you postponed any technical activities before or during the migration to microservices?} Please report the postponed activities (or group of activities) before and during the migration on the Post-it notes.

At the end of the session, the moderator presented the results obtained from the SonarQube analysis and asked for feedback to the developers.

For Q1 and Q4, each participant reported their answers on Post-it notes and attached them to a whiteboard. 
Then, each participant described their answer and grouped it with similar ones based on the group discussion. 

For Q2 and Q4, after the participants had agreed on the qualities and postponed activities, the moderator asked them to associate them with one of the ISO/IEC 25010~\cite{ISOIEC2010} quality characteristics (Functionality, Performance/Efficiency, Compatibility, Usability, Reliability, Security, Maintainability, and Portability). The moderator highlighted seven areas on a large whiteboard and asked the participants to discuss to which category each activity belongs to. 

\subsection{Data Analysis}
\label{DataAnalysis}


In order to answer RQ1, we analyzed the growth of the total TD in minutes before and after migrating to microservices. We compared the rate of growth of the TD, analyzing the overall TD (sum of TD\_M, TD\_R, and TD\_S) and all the three TD issue types independently, applying linear regression to the data before and after migrating to microservices. The growth rates of the two regression lines were compared by inspecting the slope coefficients of the lines. The $R^2$ values for the regression lines were also determined. 
We complemented the results of RQ1 obtained by the automated TD analysis by analyzing the answers of the focus group to Q2 and Q3. The results were analyzed by comparing the positive and negative answers (\rs{yes/no}).

For RQ2, we determined the relative proportion of each TD issue type in the total TD of each commit. Then, we determined whether the relative distribution of TD issue types was statistically \rs{significant} before and after migrating to microservices by applying the Mann-Whitney test. This is a non-parametric test for which the null hypothesis is that the distributions for both tested groups are identical and thus there is an even probability that a random sample from one group is larger than a random sample from the other group~\cite{mcknight2010mann}. The results are considered statistically significant if the p-value is smaller than 0.01. We calculated the Mann-Whitney test using the default value of python’s scipy package which ''computes p-value half the size of the ‘two-sided’ p-value and a different U statistic.''

Moreover, in order to determine the magnitude of the measured differences, we used Cliff's Delta, which is a non-parametric effect size test for ordinal data. The results were interpreted using guidelines provided by Grissom and Kim~\cite{grissom2005effect}. The effect size is small if $0.100 \leq |d| < 0.330$, medium if $0.330 \leq |d| < 0.474$, and large if $|d| > 0.474$.

Last, we analyzed the responses from the focus group to Q4 in order to complement the results of the aforementioned analysis. 
Since the coding regarding the different software quality characteristics was performed by the participants, we only analyzed the results using descriptive statistics.

\section{Results}
\label{Results}
In this Section, we present the main outcomes of our study.

\rss{
To answer our RQs, we first analyzed the data collected with SonarQube (RQ$_{1.1}$, RQ$_{2.1}$) and then we conducted the }focus group to provide insights into the reasons for the evolution of TD (RQ$_{1.
2}$, RQ$_{2.2}$). 

\rss{ As for the SonarQube analysis, the data we used stems from two separate time frames: before the migration (23\textsuperscript{rd} August 2014 - 30\textsuperscript{th} April 2016) and after the migration (3\textsuperscript{rd} January 2017 - 20\textsuperscript{th} September 2018). The time period between May and December 2016 is left out, as during that time the monolithic system was not actively developed. 
}

\rss{As for the focus group, }  eleven members of the company participated: four developers from one team, five developers from another team, the software architect, and the project manager. 
The focus group study lasted for two hours.

\subsection{RQ$_1$: Is the TD of a monolithic system growing with the same trend as a microservices-based system?}

\begin{figure}[h]
    \centering
    \includegraphics[height=0.4\textheight,trim={{0.1cm} {0.1cm} {0.1cm} {0.1cm}},clip]{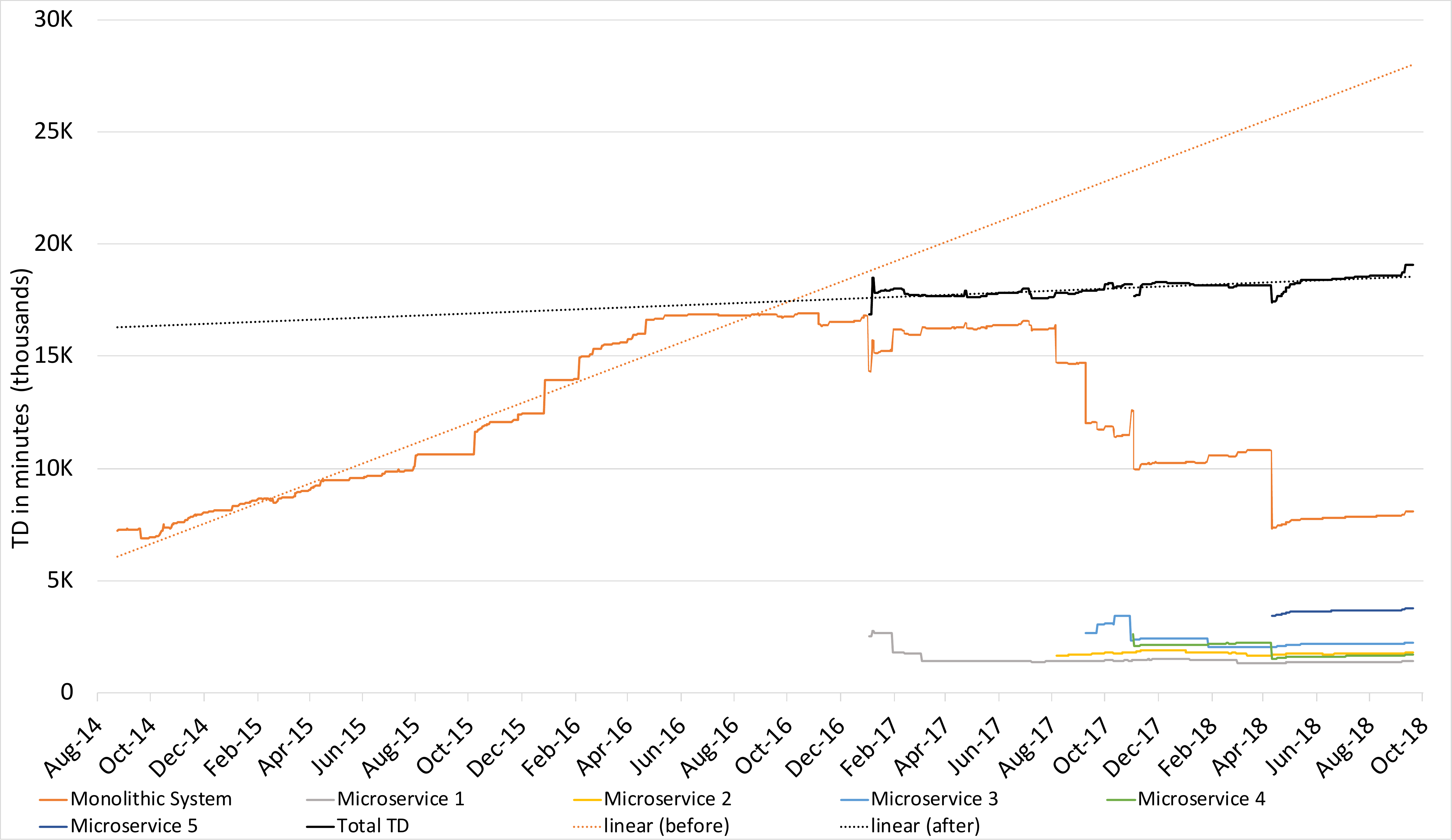}
     \caption {Overall TD evolution of the five microservices and the monolithic system (TD\_M+TD\_R+TD\_S)}
    \label{fig:OverallEvolution}
\end{figure}

Considering the trend of TD \rss{analyzed by SonarQube} (\textbf{RQ$_{1.1}$}), the overall TD (sum of TD\_M, TD\_R, and TD\_S ) of the monolithic system before the migration was lower than the overall TD right after the migration (Figure~\ref{fig:OverallEvolution}). Immediately after the introduction of a new microservice, the sum of the TD of the monolithic system and all the microservices grew faster compared to the growth of the TD before the migration. 

 As soon as a microservice became stable, the TD decreased significantly and started growing with a lower trend compared \rs{to} the growth of the monolithic system before the migration. 
Once the TD was stabilized after the extraction of a feature from the monolithic system as a new microservice, the TD trend grew much slower than before the migration. 

It is worth noting that, in the five microservices we monitored, TD commonly increases immediately after the creation of a new microservice and decreases later on. 

\rss{Figure~\ref{fig:OverallEvolution}  shows the overall TD evolution over time and the linear regression lines fitted to the data before and after the migration to microservices. The slope coefficients for the regression line and its $R^2$ value is reported in Table~\ref{tab:regression}}. The table shows that the slope coefficient for the overall TD dropped significantly after migrating to microservices, and that the coefficient after migration was 89.76\% smaller than when using a monolithic system. This implies a remarkable drop in the overall TD growth rate.

\begin{figure}[h]
    \centering
    \includegraphics[height=0.4\textheight,trim={{0.1cm} {0.1cm} {0.1cm} {0.1cm}},clip]{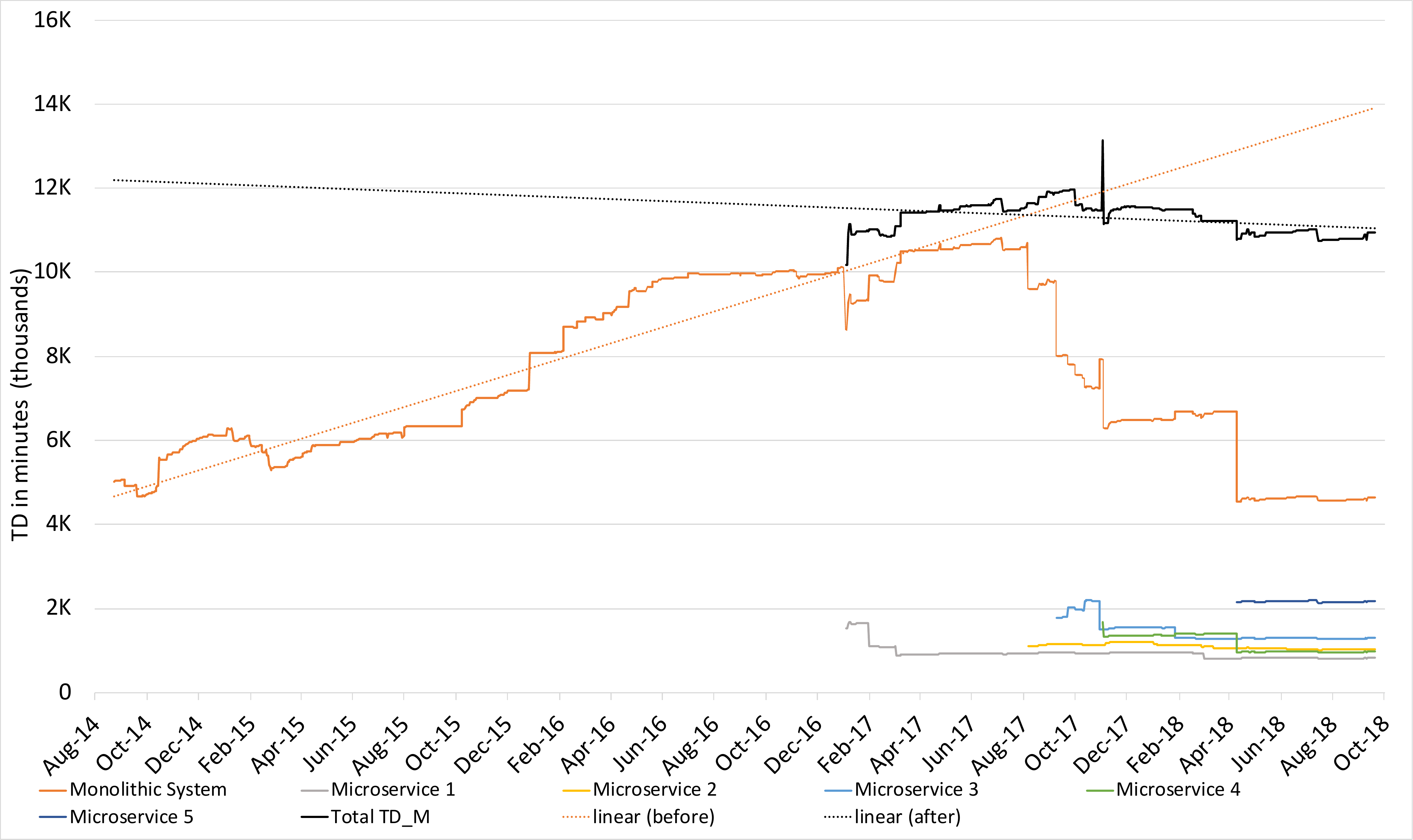}
     \caption {TD\_M evolution of the five microservices and the monolithic system}
    \label{fig:TDMEvolution}
\end{figure}

\begin{table} [H]
\centering
\small
\caption{TD slope coefficients before and after the migration to microservices.}
\label{tab:regression}
\begin{tabular}{c|c|c|c|c|c}
        \hline
        & & \multicolumn{3}{|c}{\textbf{Type}} \\
         & & TD\_M & TD\_R & TD\_S & TD \\
         \hline
        \multirow{3}{*}{\textbf{Before migration}} & Slope coeff. & 6.21 & 1.11 & 7.38 & 14.71 \\
        & $R^2$ & 0.81 & 0.55 & 0.97 & 0.93 \\
        \hline
        \multirow{3}{*}{\textbf{After migration}} & Slope coeff. & -0.77 & -0.54 & 2.81 & 1.51 \\
        & $R^2$ & 0.16 & 0.48 & 0.60 & 0.64 \\
        \hline
\end{tabular}
\end{table}

\begin{table} [H]
\centering
\small
\caption{The descriptive statistics for TD types before and after the migration and the results from the Mann-Whitney and Cliff's Delta tests.}
\label{tab:manwhitney}
\begin{tabular}{c|c|c|c|c}
        \hline
        & & \multicolumn{3}{|c}{\textbf{Type}} \\
         & & TD\_M & TD\_R & TD\_S \\
         \hline
        \multirow{3}{*}{\textbf{Before migration}} & mean & 0.95 & 0.012 & 0.035 \\
        & median & 0.95 & 0.012 & 0.037 \\
        & stdev  & 0.01 & 0.003 & 0.005 \\
        \hline
        \multirow{3}{*}{\textbf{After migration}} & mean  & 0.97 & 0.004 & 0.028 \\
        & median & 0.97 & 0.005 & 0.027 \\
        & stdev  & 0.01 & 0.001 & 0.005 \\
        \hline
        \hline
         \multirow{2}{*}{\textbf{Mann-Whitney}} & U & 18,066 & 0 & 57,269 \\
         & p-value & 0 & 0 & 0 \\ 
         \hline
         \multirow{2}{*}{\textbf{Cliff's Delta}} & d & -0.91 & 1 & 0.70 \\
         & CI & -0.93 - -0.88 & 1.00 - 1.00 & 0.66 - 0.74 \\
         \hline
\end{tabular}
\end{table}

\rss{The developers' perceived TD also confirms this trend (\textbf{RQ$_{1.2}$}}). 

\rss{During the focus group,} the developers confirmed that during the first months after the introduction of a new microservice, several activities were postponed (Q4). In particular, the participants reported that a total of 29 activities were postponed during the development of the monolithic system, while 30 were postponed during the development of the microservices (Table~\ref{tab:reasons}). 
This was done because they wanted to prioritize the delivery of the system in production. 

\rss{Developers explained the vertical drops in the SonarQube TD, even beyond the points of introduction of new microservices, due to the process they adopted to create the microservices. Microservices were initially created by copying the code from the monolithic system into the microservice. Then, developers started to integrate and refactor the code, with the result of obtaining more clean code and reducing TD. This process took a different time for each microservice, and this might  explain the different  delays on the  reduction of TD. 
}

\rss{The participants reported that the increased amount of perceived TD \underline{during the introduction of new microservices}, was mainly due to two reasons:}
\begin{itemize}
    \item The delivery of the system in production had the highest priority, and sometimes required to make temporary sub-optimal choices, to deliver on time.  
    \item Need for writing the code connecting the monolithic system to the microservices. The rationale behind this was that the development of a new microservice involves duplicating the existing service and adopting a large number of temporary solutions before the microservice becomes stable.  
\end{itemize}

Compared to the monolithic system, all the participants agreed that the activities postponed in the microservices, even if they will accrue TD, will be much easier to implement in the microservices than they would have been in the monolithic system. They named the limited size of the microservices as the main reason for this.

Regarding the amount of perceived TD \underline{after the migration} (Focus Group - Q3), the developers, the project manager, and the software architect had different opinions.
The software architect and the project manager perceived the increase of effort \rss{in the microservices} as a negative issue, since the overall development cost increased by more than 20\% due to the cost of developing a distributed system. The increased overhead is in line with previous research~\cite{SOLDANI2018215}\cite{Taibi2017}.

\rss{The software architect was also concerned about the architectural decisions postponed due to the urgent need of releasing new features or fixing bugs. However, both the project manager and the software architect felt that the TD in the microservice-based system is lower and also grows slower than in the monolithic system. }

\rss{Developers also confirmed that TD decreased after the introduction of the microservice and accumulates slower compared to the legacy monolithic system. Moreover, they also reported being less under pressure after each microservice became stable, compared to the work pressure they felt during the development of the monolithic system. }


Below are several examples of postponed technical decisions. For example, in two microservices, the team kept the old SQL database, even though they had planned to migrate to a NoSQL database. Another postponed decision regarded the outdated libraries in the monolithic system. Upgrading the libraries to newer versions would have required several changes in the code, and the team decided to postpone doing this due to time constraints. 
The third important postponed activity was the refactoring of the code extracted for microservice 5. The developers had planned to deeply refactor the code extracted from the monolithic system before using it in production. However, the refactoring had to be postponed because of an urgent new feature request impacting microservice 5. The team reported that they had the choice of either implementing the new feature in the monolithic system and then applying all the changes to the almost migrated microservice, or implementing the new feature directly in the new microservice.  
Yet another important postponed decision was the development of an API-Gateway. The company used the RabbitMQ message bus as an API-Gateway, but they are planning to migrate to a proper API-Gateway in the future.  
Overall, despite all the voluntarily postponed activities, all the participants perceived the overall TD as decreased.

\roundedbox{
\textbf{RQ$_1$ Summary}: \textit{
TD detected by SonarQube grew ~90\% slower in the microservice-based system. After the initial introduction of each microservice, TD grew for a limited period of time, because of the legacy code was reused as it was. When the code of the microservice was completely refactored and stabilized, TD decreased and started growing linearly with a growing trend much lower than in the monolithic system. 
TD perceived by the development team also follows the same trends.
}}


\subsection{RQ$_2$: How the different types of TD change after the migration to microservices?}

\begin{figure}[h]
    \centering
    \includegraphics[width=0.9\textwidth,trim={{0.1cm} {0.1cm} {0.1cm} {0.1cm}},clip]{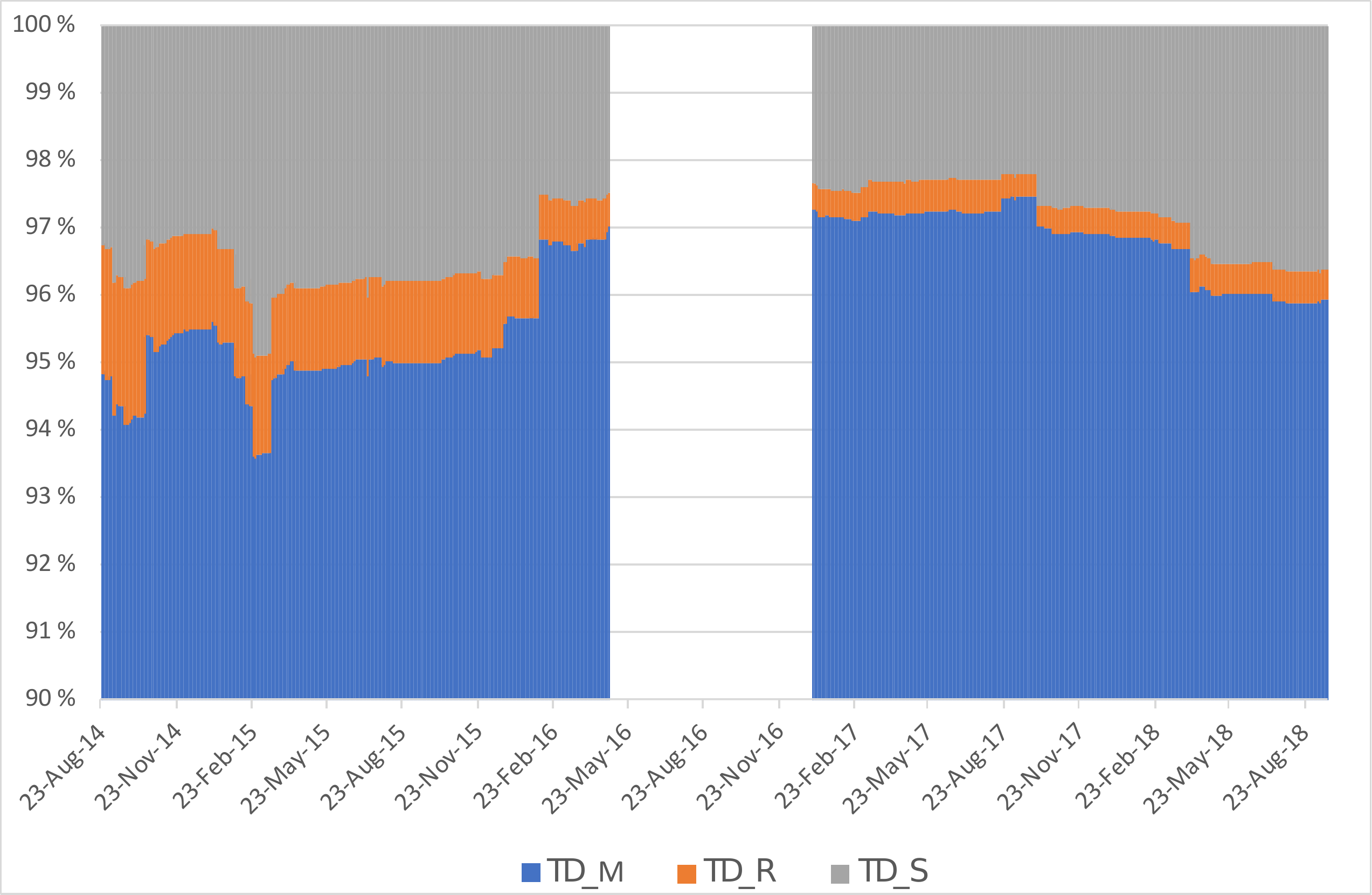}
    \caption{Relative distribution of TD issue types.}
    \label{fig:relative_types}
\end{figure}

As for distribution of the different SonarQube TD types \rss{(\textbf{RQ$_{2.1}$} -  Figure~\ref{fig:relative_types})}, 
in order to determine whether the change in the relative distributions between the different TD issue types 
was statistically significant, we applied the Mann-Whitney test for each type, reporting the results in Table~\ref{tab:manwhitney}. As the p-values are smaller than 0.01 for all types, we conclude that the relative distributions of TD issue types did change after migration to microservices. The results from the Cliff's Delta test are also presented in Table~\ref{tab:manwhitney}. All measured values of $|d|$ were greater than 0.474 and thus the effect size is considered large for all types. Also, the confidence intervals (CI) for $d$ using 0.99 confidence are greater than 0.474.


Looking at the distinct TD types (Figure~\ref{fig:relative_types}), a vast difference can be noticed as well after the migration to microservices. In particular, TD\_M and TD\_R had a negative coefficient after the migration, indicating a decrease in TD after the adoption of microservices. Also, the coefficient of TD\_S was reduced, resulting in a much slower TD growth rate.

Since the vast majority of business processes are still in the monolithic system after the migration, we expect the TD\_M of the whole system to be lower than the TD\_M of the monolithic system after all of the business processes have been migrated.

\begin{figure}[h]
    \centering
    \includegraphics[height=0.4\textheight,trim={{0.1cm} {0.1cm} {0.1cm} {0.1cm}},clip]{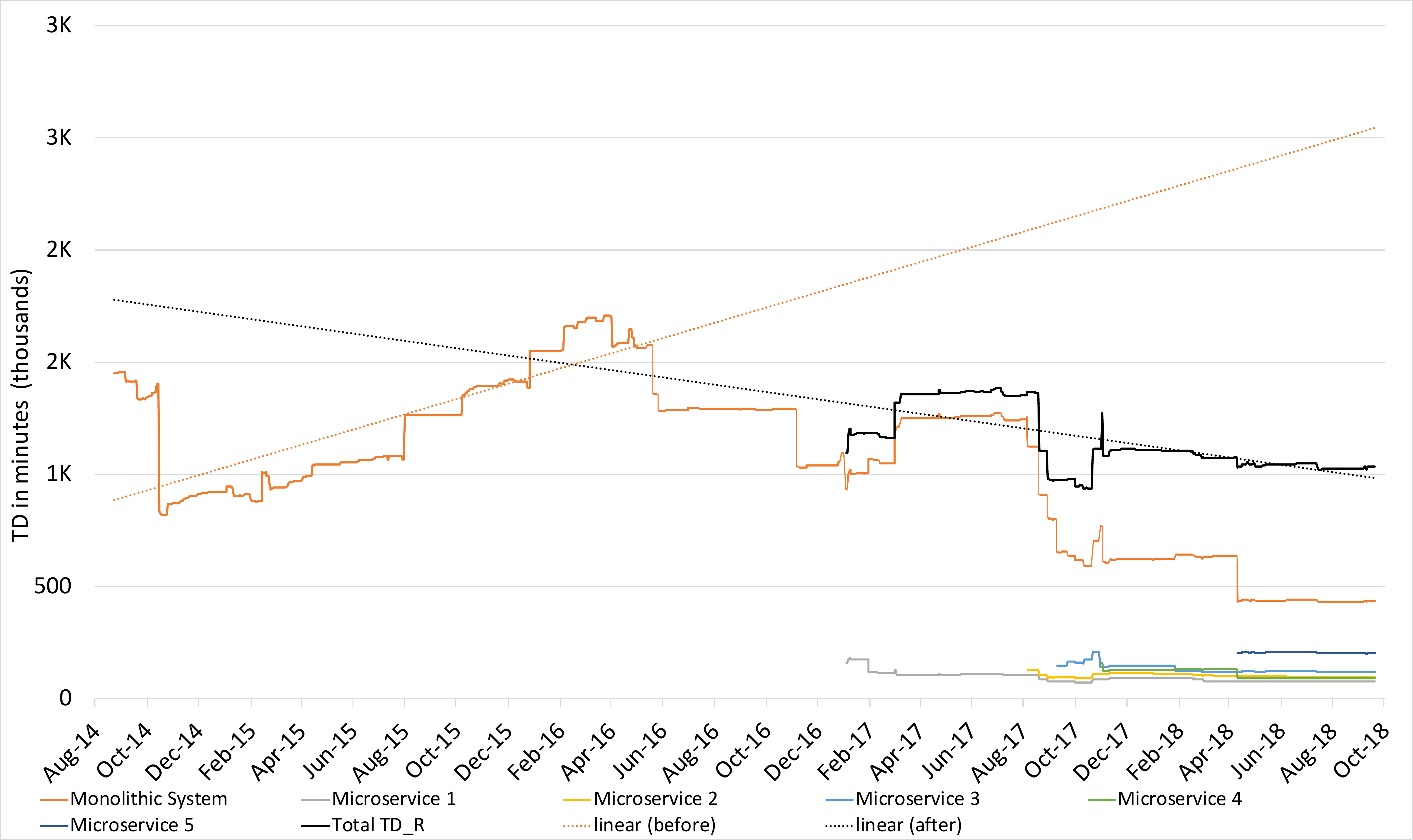}
     \caption {TD$\_$R  evolution of the five microservices and the monolithic system}
    \label{fig:ReliabilityEvolution}
\end{figure}

\begin{figure}[h]
    \centering
    \includegraphics[height=0.4\textheight,trim={{0.1cm} {0.1cm} {0.1cm} {0.1cm}},clip]{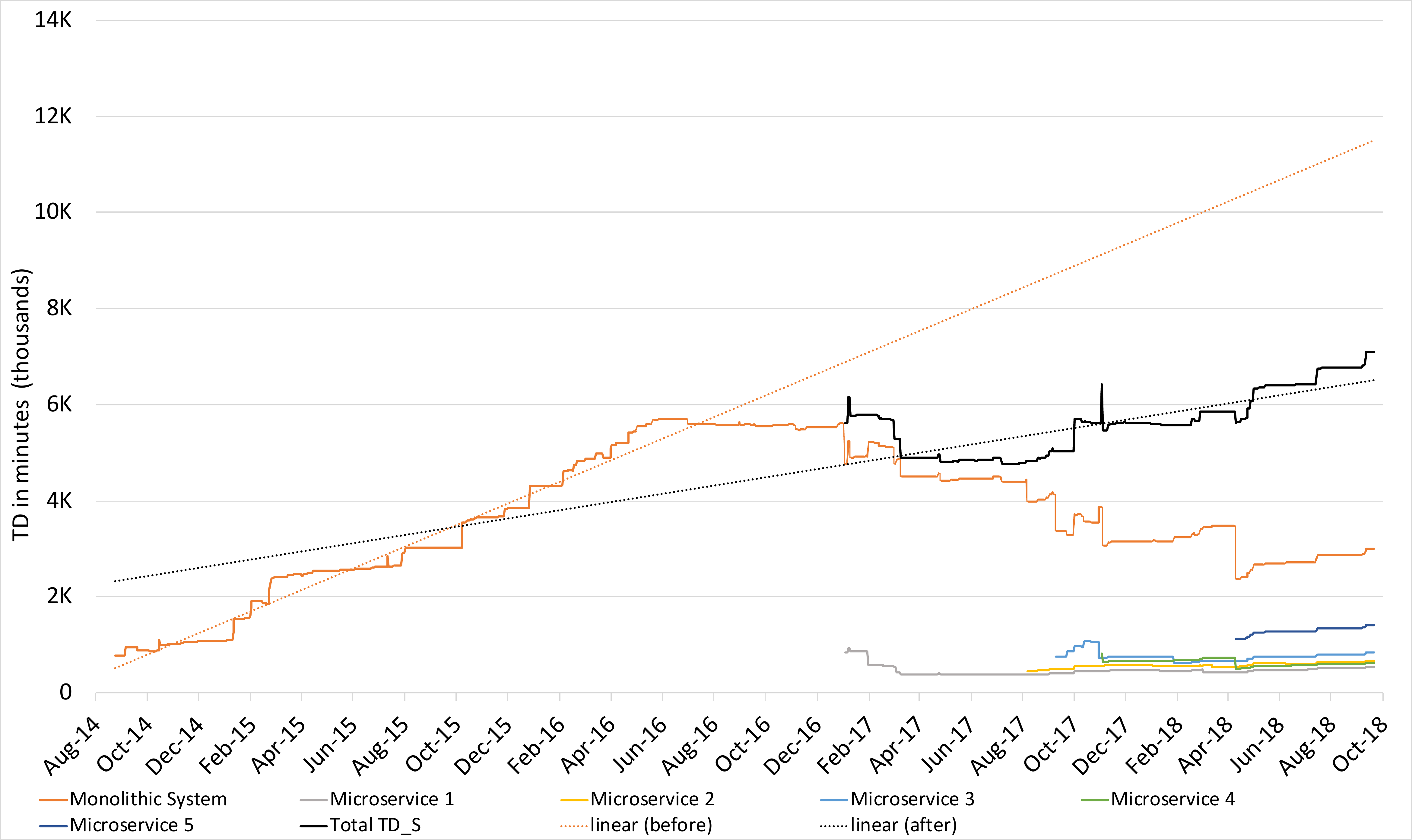}
     \caption {TD$\_$S evolution of the five microservices and the monolithic system}
    \label{fig:VulnerabilityEvolution}
\end{figure}

Figures~\ref{fig:TDMEvolution},~\ref{fig:ReliabilityEvolution}, and~\ref{fig:VulnerabilityEvolution} show the linear regression lines fitted to the data before and after the migration to microservices together with the individual TD related to vulnerability, reliability, and maintainability, respectively. The slope coefficients for the regression lines and their $R^2$ values are reported in Table~\ref{tab:regression}.

\rss{As in \textbf{RQ$_1$}}, the focus group enabled us to confirm the trend reported by SonarQube and get more insights on the different qualities affected by the migration \rss{(\textbf{RQ$_{2.1}$})}.  

As for the overall code quality (Focus Group - Q2), all of the participants confirmed that the code in the microservices is easier to read and modify. They also selected the created microservices as the components with the highest quality. They did so even though the migration process did not involve a complete rewrite of the code, as pointed out by the developers. Instead, they refactored only parts of the code and largely reused the existing code.

The participants assigned the postponed activities to four groups based on their quality characteristics (Focus Group Q4): Performance/Efficiency, Reliability, Security, and Maintainability. \rss{No activity was assigned to the ISO/IEC 25010~\cite{ISOIEC2010} quality characteristics: Functionality, Compatibility, and Usability. } Each activity was assigned to only one group. Details on the number of postponed activities are reported in Table~\ref{tab:reasons}.

It is worth noting that three out of four quality characteristics correspond to the TD categories proposed by SonarQube. Moreover, even though the results are only related to the number of postponed activities and not to the time needed to refactor them, we can still see an increased number of postponed activities related to security characteristics. 

The developers reported that during the development of the microservices, they paid a lot of attention to the future \textbf{maintainability} of the system. A reduction in the number of postponed activities regarding maintainability-related aspects was therefore expected. This result is also in line with the results obtained from SonarQube (RQ$_{2.1}$), where the TD\_M trend is inverted compared to the trend in the monolithic system. After the introduction of each microservice, the TD\_M tended to decrease instead of increasing as before. 

Developers also reported that the migration had an impact on testing the system. They reported that unit testing is easier in the microservices. However, integration testing was considered much more complicated, mainly because it is more complex to replicate the whole system locally on a developer's machine. No differences were reported for end-to-end testing. All the old Selenium tests\footnote{Selenium - Web Browser Automation. https://www.seleniumhq.org Last access: August 2019} that were performed on the graphical user interface were executed without changes. 

Considering security-related activities, the developers reported that the migration to microservices forced them to postpone more activities compared to when they developed the monolithic system. One of the main issues experienced by the developers was that there was a single sign-on approach between applications. The developers initially extended the authentication system of the monolithic system to provide authorizations to the individual services instead of implementing a system based on OAuth 2.0\footnote{OAuth 2. https://oauth.net/2/}, as planned by the software architect. The postponement of this activity was due to the time pressure for delivering the new updates for the next tax year. 

As for performance- and efficiency-related activities, the developers only postponed one activity during the development of the microservices. They reported that the performance had generally not been very important in the past, as the system never had important performance issues. The only exception was uploading of the PDF invoices, which is a task a tax accountant usually performs once a year in batch. Invoices need to be processed, sent to the Italian Ministry of Economy and Finance, and then stored permanently. The software architect and the development team planned to implement this process with a serverless function to enhance the performance and avoid over-sizing the whole infrastructure for an activity performed only once a year. However, time constraints forced the developers to postpone this activity to the next year as well. 

Regarding reliability, the developers invested a lot of effort into getting a reliable system. However, they had to postpone some activities because of time pressure and technical reasons.
Because of the NDA, we are not allowed to describe all the postponed activities.

\begin{table}[H]
\centering
\caption{Activities postponed before and during the migration to microservices }
\label{tab:reasons}
\footnotesize
\begin{tabular}{l|c|c}
\hline
& \multicolumn{1}{c|}{\textbf{Before the Migration}} & \multicolumn{1}{l}{\textbf{During the Migration}}  \\ \hline
Performance/Efficiency & 3                                                 & 1                                                  \\
Reliability            & 5                                                 & 2                                                  \\
Security               & 10                                                & 18                                                 \\
Maintainability        & 11                                                & 9    \\
\hline
\textbf{Total}          & 29                                               & 30 \\ 
\hline
\end{tabular}
\end{table}

\roundedbox{
\textbf{RQ$_2$ Summary}: \textit{
All the three types of TD decreased  significantly after the microservices got stable, with an important drop rate in T\_M and TD\_R.
This trend is also confirmed by developers, that they paid attention to the different qualities of the microservices. }}

\section{Discussion}
\label{Discussion}

The migration to microservices is a non-trivial task that requires deep re-engineering of the whole system. This heavily impacts on the whole project cost, but should also facilitate maintenance in the long run. 

Our case company highlighted that if the system had remained monolithic, they might have missed the deadlines of the annual updates of the system. This was caused by the constant grow of TD, as each year it took longer to adapt the system to the new tax rules. Therefore, despite the increased overall development costs, the migration was considered beneficial. The reasons for the extra costs were manifold. First of all, the developers had to deal with a new system architecture. They also had to consider various aspects, such as enabling the legacy system to communicate via Enterprise Service Bus with the microservices, dealing with authentication issues, as well as with process-related issues such as the introduction of the DevOps culture.

\rss{Our case study on the migration to microservices confirmed our expectations. As expected, the TD in the monolithic system was growing with a higher rate than with microservices. 
This is confirmed by the significant decrease of the maintenance predictors proposed by SonarQube, and thus a reduction of the remediation effort related to "code smells", and by the perception of the developers. Despite developers postponed a similar number of activities during the development of the monolithic and of the microservice-based system, they perceived a lower amount of TD, since they consider the activities postponed in the microservice easier to implement and less trivial then in the monolithic system. }

The vast majority of postponed activities were related to quality aspects, except for unit and integration tests. The most important type of postponed activities were those related to architectural decisions, mainly because they were not aware of common patterns and anti-patterns. As an example, the developers connected all microservices point-to-point, instead of using a message bus during the first period, and then they had to refactor the systems, introducing the message bus later. 

The focus group enabled us to get more insights on the benefits and issues the company faced migrating to microservices. 

The main benefits were decreased bug-fixing time and a large increase in the system's understandability. These were also the main initial expectations the company had had for the migration. Other minor benefits were the reduced need for synchronization between teams and the possibility to deploy new features without re-compiling and re-deploying the whole system. 

However, regarding the main issues, the project manager and the software architect highlighted higher development costs, possibly because the developers used microservices for the first time. The developers highlighted the complexity of connecting the different microservices to each other or to the monolithic system. The monolithic system had the advantage of local calls, while with microservices, they had to rely on a distributed system. The extra cost was considered positive, as it included some activities that had not been performed in the past or that had been postponed. With the introduction of microservices, the company also introduced continuous delivery and increased the test coverage of the system in order to have higher confidence in the automatic builds. 
\rss{Other important issues were related to the availability of a microservice template, that in the case of the first microservice was not perfectly suitable, and required some extra work. }
The initial increase of TD after the introduction of each microservice was probably due to the lack of a service template at the beginning. The second microservice was not easy to develop, but it was easier than developing the first one because of the availability of the template developed for the first microservice. We recommend other companies investing more effort into the definition of a set of service templates, as this will dramatically ease the development of new microservices in the future.


Another important issue was related to the postponement of architectural decisions. Continuous architecture principles recommend postponing architectural decisions until they are really needed~\cite{ErderPierre16}. However, we observed that implementing postponed architectural decisions requires significantly more effort than it would have initially required. An example was the usage of the lightweight message bus (RabbitMQ) as API-Gateway instead of the using a proper API-Gateway. The SME of our case study is still using RabbitMQ as their API-Gateway. However, they are aware that implementing a proper API-Gateway at the beginning would have cost much less than migrating from RabbitMQ after two years.

\rff{As for the results obtained by the SonarQube analysis, developers confirmed their interest in the measurement, and especially in security and reliability rules. The team analyzed the most recurring violations in their monolithic and in the microservices, realizing that rules commonly violated in the monolithic system were not commonly violated in the microservices. The reason might be that developers are now more trained on quality aspects than in the past, and that during the development of the microservices they also paid a special attention on writing clean code. }

\rff{Developers did not agree on the importance, and especially on the severity, of several rules considered harmful by SonarQube. As an example,  they did not consider harmful the rule ''Generic exceptions should never be thrown''. SonarQube reports that using such generic exceptions as ''Error'', ''RuntimeException'', ''Throwable'', and ''Exception'' prevents calling methods from handling true, system-generated exceptions differently than application-generated errors. This rule is classified by SonarQube as a ''Code Smell'' (rules that decrease code maintainability) and of major severity. 
However, developers appreciated the availability of other rules, that in their opinion are more severe than what is reported by SonarQube. 
As an example, they consider rules related to unused variables but especially to unused classes and methods, that enable them to understand if the method or class should be removed or if it should have been used by another method but for some unexpected reason it is not used. 
Developers also rated the importance of the security rules high, as they enable them to spot security vulnerabilities they were not able to manually identify. 
}

\rff{As a result of the focus group, the company decided to adopt SonarQube in production and to integrate it into their CD/CI pipeline. The next step will be the identification of the actually harmful rules. For this purpose, in future works, we will support them investigating the fault- and change-prone rules combining their experience with the results of the actual change- and fault-proneness of the rules detected on their source code using a similar approach to the one we adopted on open-source projects~\cite{SaarimakiTechDebt2019}\cite{LenarduzziSANER2019}\cite{LenarduzziSomeEffect2020}}. 

Based on the the experience gained in this work, we recommend companies to first evaluate the need to migrate to microservices~\cite{Auer2020} and to carefully investigate possible migration opions~\cite{TaibiSysta2019}. Then, once they started to develop the microservice-based system, to continuously apply quality monitoring approaches in their development process, monitoring  the technical debt and code quality~\cite{JanesMOLS2017}\cite{ECISME} but also technical debt at the architectural levels, so as to avoid the presence of microservice anti-patterns~\cite{Taibi2020} and bad smells~\cite{Pigazzini2020}. 

\section{Threats to Validity}
\label{Threats}
In this Section, we will introduce the threats to validity following the structure suggested by Yin~\cite{YinCaseStudies2009}, reporting construct validity, internal validity, external validity, and reliability. Moreover, we will also debate the different tactics adopted to mitigate them.

\textit{Construct Validity} concerns the identification of the measures adopted for the concepts studied in this work. 
We analyzed the TD using the model provided by SonarQube. Therefore, different tools and approaches might have provided different results. 
We are aware that other types of TD, such as requirements TD or architectural TD, can heavily impact the TD of a system. We are aware that the accuracy of the TD detected by SonarQube might be biased ~\cite{LenarduzziEuromicro2019}.  However, SonarQube was the only tool that we were allowed to use at the company. 
We are also aware that important postponed activities could have created a large amount of TD. We mitigated this threat by performing the focus group and by asking the teams to discuss other possible types of TD. A more thorough discussion on the amount of TD that these postponed activities will generate will be part of our future work. 

Threats to \textit{Internal Validity} concern factors that could have influenced the obtained results. 
The postponed activities were collected using a group discussion. It is possible that some developers did not want to publicly expose some activities they had postponed.  

Threats to \textit{External Validity} concern the generalization of the obtained results. 
The results of this paper are based on the monitoring of the development process of a single company. The results could be slightly different in other companies. However, based on previous studies on microservices, the developers confirmed that microservices increase maintainability, code readability, and system understandability~\cite{Taibi2017}.  Therefore, we expect that other systems could also benefit from a decrease of the TD when migrating to microservices.
\rss{The microservices we investigated were structurally and functionally similar. Therefore, systems implementing microservices with different structures or functionalities might experience different results.}

Threats to \textit{Reliability} refer to the correctness of the conclusions reached in the study. 
This study was a preliminary study, and therefore we applied simple statistical techniques to compare the trends of the TD before and after the migration. The results of the statistical technique applied are also confirmed by Figure~\ref{fig:OverallEvolution}. We are aware that more accurate statistical techniques for time series could have provided a more accurate estimate of the difference of the slopes. However, we do not expect that different statistical techniques would provide a contradictory result.

\section{Conclusion}
\label{Conclusion}

In this work, we compared the technical debt (TD) before and after the migration to microservices of a twelve-year-old software project (280K Lines of Code) developed by an Italian SME. This is one of the first studies investigating Technical Debt in SMEs adopting Microservices~\cite{LenarduzziESEM2019}.

We conducted a case study \rss{to investigate the different perspectives of TD during the migration of microservices, including the TD detected at code level by SonarQube and perceived by the developers}

For this purpose, we analyzed \rss{with SonarQube} the code TD of a system under development (two years before and two years after the migration). Then, we conducted a focus group to analyze the developers' perceived TD and the related postponed activities in depth and to get more insights into the results. 

The first result revealed that TD grew ~90\% slower after the development of microservices. After the initial introduction of each microservice, TD grew for a limited period of time, mainly because of the new development activities. When the code of the microservice stabilized, TD decreased and started growing linearly, with a growing trend much lower than for the monolithic system. 

Unexpectedly, when comparing the distributions of TD issue types before and after the introduction of microservices, important and statistically significant differences emerged. The proportion of SonarQube issues classified as bugs and security vulnerabilities decreased, while code smells (maintainability issues) increased. Since microservices are supposed to facilitate software maintenance, we expected a reduction of code smells.

The developers confirmed the overall results, perceiving reduced maintenance complexity and increased velocity. 
The overall development effort increased after the introduction of microservices because of the extra effort due to the re-development of the system, the connection of the legacy system to the new microservices, introduction of a distributed authentication mechanism, and many other activities not previously considered. However, the manager confirmed that the increased velocity and team freedom compensated for the required extra effort.  For the company, it was especially important that the migration allowed them to remain on the market, releasing the annual tax rules update required by the government on time. 

Future work will include an investigation of the impact of other types of TD during the migration to microservices. We aim to analyze the same projects using tools for detecting architectural smells. Moreover, we aim to investigate TD due to temporary architectural decisions. 
Our next goal is to understand how long different activities could be postponed before the benefit of postponing an activity is canceled out by the increased effort needed to refactor it. As an example, if an activity has an interest equal to zero (i.e., if the development/refactoring effort does not increase if postponed), it can be postponed until it is needed, whereas if an activity has a monthly interest of 10\% (i.e., 10\% extra interest per month), it should be refactored as soon as possible before becoming too expensive. Last, but not least, we are also planning to investigate the evolution of TD in companies adopting serverless functions~\cite{Nupponen2020} for the development of microservices-based systems, so as to prevent anti-patterns and to keep TD under control~\cite{TaibiIEEE2020}.

As the company finally decided to adopt SonarQube in production, another future work will focus on the identification of harmful SonarQube rules and therefore, an accurate definition of the quality model to be used. 
\section*{Acknowledgements}
This research was partially supported by the ''SHAPIC'' grant awarded by the Ulla Tuominen Foundation (Finland)

\section*{References}
\bibliographystyle{model1-num-names}
\bibliography{sample}
\begin{appendices}
\section{The migration process adopted by the case company}
\label{app:migration_process}

In this Appendix, we summarize the process adopted by the case company to migrate their monolithic system to microservice. 

 The company started discussing the migration to microservices in April 2016. 
  The software architect, together with the support of two consultants with experience in migration to microservices, analyzed the feasibility and the potential usefulness of the migration, estimating costs and evaluating the availability and need of resources. 
 
They decided to migrate, despite an estimated high initial cost due to the migration overhead, as microservices often increase development velocity and enable teams to work independently, therefore reducing the hassle during the very short time they have for implementing the new features every year. Moreover, they decided to hire two consultants with experience in the development of microservices and experience in decomposing monolithic systems into microservices.
 
 The company froze the development of the system between May and December 2016, implementing only critical bug fixes. During this time frame, the company sent the developers to training courses on microservices, and the software architect, with the support of the two consultants, started the analysis of the monolithic system and planned the migration.

 In January 2017, a team composed by four internal developers, and the two consultants started implementing the first microservice. The consultants had more than five years of experience with microservices. \rff{The four internal members were experienced developers (two with more than 20 years of experience in Java, and two with five years of experience in Java), but with no experience on developing microservice, except from the experience they had during the training course}.

 The migration process was implemented in three main steps. 
    \begin{itemize}
      \item[Step 1] \textit{Identification of decomposition options, architectural guidelines, and migration plan.} The software architect and the consultants sliced the system into independent microservices. First they analyzed the internal dependencies with Structure 101\footnote{Structure101 Software Architecture Environment - http://www.structure101.com}. Then they identified decomposition options based on the different services corresponding to the business capabilities and proposed them together with a set of architectural guidelines that all the extracted microservices should follow. As an example, they decided that microservices should not communicate directly with each other but should use the publish–subscribe pattern to communicate through the RabbitMQ message bus\footnote{RabbitMQ Message Broker. Online: https://www.rabbitmq.com}. Moreover, they also decided to temporarily use RabbitMQ as API-Gateway. Finally, they involved all of the team members in the prioritization of the microservices, discussing which services should be developed with the highest priority. The priorities were assigned based on different criteria. In some cases, services had high priority because they had a lot of bugs, and the re-implementation as a microservice would enable the developers to completely re-develop from scratch and thus to fix the issues. In other cases, microservices were prioritized based on their business priority. The only major constraint imposed of the company was that during the period from December to April, only one service could be migrated, as the highest priority had to be on adapting the system to the new tax laws.  
      \item[Step 2] \textit{Implementation of the first microservice}. The company decided to implement some of the new regulations in the first microservice, instead of implementing them in the monolithic system and then migrating them later. The first microservice was based on a low-risk component, which would have made it possible to move all changes to the monolithic system if major issues had arisen.  \\
            Before the implementation of the first microservice, the consultants provided a skeleton of a sample microservice and supported the company in setting up a continuous delivery pipeline using Gitlab-CI\footnote{https://about.gitlab.com/product/continuous-integration/}. 
      
    \item[Step 3] \textit{Implementation of the other microservices}. Based on the results of the implementation of the first microservice, the other teams gradually started the implementation of other microservices, based on the plan proposed in Step 1. 
  \end{itemize}

\section{Study Protocol}

As reported by Runenson and H\"{o}st~\cite{Runeson2009}, it is important to report a case study protocol for case studies. For this purpose, in this Section, we report protocol we adopted, following the guidelines they recommend~\cite{maimbo2005}.

The purpose of the protocol is to define the detailed procedures for collection and analysis of the raw data, enabling other researchers to replicate this study in other contexts.


This study was based on two sub-studies.
\begin{itemize}
    \item Sub-Study 1:  \textit{Automated TD analysis}: In this step, we  collected data on code TD in system,  before and during the migration.
    \item Sub-Study 2: \textit{Focus group } to understand the usefulness of the TD analysis and gather more qualitative insights
\end{itemize}

In the next Sections, we describe the procedure for the collection and analysis of the two sub-studies. 

\subsection{Sub-Study 1: Automated TD Analysis}

In this Step, we analyzed the TD of the project using SonarQube SonarQube\footnote{https://www.sonarqube.org} (version 7.0). 


\subsubsection{Procedures}
For the automatic TD analysis, we need access to the source code of the company, and the possibility to execute SonarQube on the code. 
For this purpose, one author scheduled weekly visits to the company where he was able to access to the source code, collect the data, and execute the analysis. As for NDA, all the analysis must be performed on the company servers. The company provided a desk for our researcher, and a virtual machine with 64 cores and 65GB of RAM for performing the analysis and install SonarQube. 
\subsubsection{Data Collection}
We first analyzed each commit of the monolithic project, starting from August 2014, using the SonarQube default model. Then we extracted the data from the SonarQube APIs.
\begin{itemize}
\item \textbf{SonarQube Analysis of all Commits:} We first analyzed each commit of the monolithic project, starting from August 2014, using the SonarQube default model.
For this purpose, we developed a script to checkout each commit from the git repository in chronological order, and execute the SonarQube analysis using the Sonar-Scanner for Maven\footnote{Sonar-Scanner for Maven https://docs.sonarqube.org/latest/analysis/scan/sonarscanner-for-maven/}. It is important to note that the analysis performed using the libraries in the private Maven repository, hosted in the company servers, to guarantee a successful compilation SonarQube analysis. 
\\ Then, we analyzed separately each microservices following the same approach. 

\item \textbf{Data Extraction from SonarQube}
After the monolithic project and the microservices had been analyzed with SonarQube, we extracted the data related to the projects from SonarQube API, using the \textit{''SonarQube Exporter''} version 0.4.0\footnote{https://github.com/clowee/SonarQube-Exporter/releases}. \textit{''SonarQube Exporter''} makes it possible to extract SonarQube project measures and issues in CSV format. The results provided by the SonarQube Exporter have the same structure of the \textit{Technical Debt Dataset}\cite{LenarduzziPromise2019}, with a table ''Measures'' that reports all the measures analyzed by SonarQube such as lines of code, code complexity and amount of comment lines, and the table ''Issues'' that include all the issues reported by SonarQube. 
\end{itemize}

\subsection{Sub-Study 2: Focus Group}

The purpose of the Focus Group was to  gain more qualitative insights on the results provided by SonarQube
The focus group was based on a face-to-face semi-structured interview. The goal was to discuss whether the migration to microservices had been beneficial and whether the participants had experienced any of the expected benefits.

The focus group was moderated by one of the authors. The whole focus group session was audio-recorded. To be able to do this, we obtained written consent from all the participants. The relevant parts were transcribed and then coded by the moderator of the focus group, and verified by one of the authors.

\subsubsection{Procedures}
The focus group was planned to last up to 2.5 hours. We selected a limited number of issues to be covered so that sufficient time can be allocated for the participants to comprehend the issue and have a meaningful discussion and interaction about them. 

\textbf{Selection of the Participants:} 
Participants were selected among the developers of the system. We invited all the team members involved in the development of the monolithic and microservice system, except for developers with less than one year of experience in the monolithic system.
Before the focus group session, we sent an invitation email to the potential participants, also reporting that the session would have been audio-recorded.

\textbf{Focus Group Session Conduction:} The session was moderated by our last author, that carefully took care of managing the schedule  to make sure that all main contributions can be made during the allocated time. 

The session was initiated by welcoming the participants and collecting the informed consents, where participants  agreed on collecting the data and audio-recording the session. Then, the moderator made an introduction describing the goals and ground rules of the session. 

After the introductory session, that was planned to last 10 minutes, each of the topics was  presented one after another.

\textbf{Tasks:}
The moderator acted as a chair of the session, supporting the discussion and guiding the participants thru four different tasks:

\begin{itemize}
    \item T1: Identification of benefits and issues perceived after the migration to microservices
    \begin{itemize}
        \item Write the issues on the red Post-it notes and the benefits on the yellow ones
        \item Collaboratively group the benefits and issues. Each member read their issue and benefits and stick the Post-it note with the issues on the left side of the wall and the ones with benefits on the right side of the wall. In case of similar issues, the group discussed if the issue or benefit must be merged with an existing one.         
    \end{itemize}                                                                                                   
    \item T2: Identification of perceived changes of quality                                           
     \begin{itemize}
        \item Write the qualities that changed on the Post-it notes                                                         \item Collaboratively group and discuss the qualities. The moderator created eight areas on the wall, one for each of the~ISO/IEC 25010~\cite{ISOIEC2010} quality characteristics (Functionality, Performance/Efficiency, Compatibility, Usability, Reliability, Security, Maintainability, Portability). Then, each participant red their quality and sticked the Post-it note on the whiteboard. In case of similar qualities, the group discussed if the issue or benefit must be merged with an existing one.
    \end{itemize}
    \item T3: Identification of perceived change on the TD
    \begin{itemize}
        \item   The moderator first introduced the concept of TD. Then asked the participant to discuss on their feeling on the TD, if it was increased or not.
    \end{itemize}
    \item T4: Identification of postponed ~activities before or during the migration to microservices 
     \begin{itemize}
        \item   Write the activities you postponed before the migration on the red Post-it notes and the ones you postponed during the development of the microservices on the yellow ones.                                 
        \item Collaboratively group and discuss the activities and associate them with the quality characteristics reported in T2                                                                                                               
    \end{itemize}                                                                                                             \item General Feedback on the SonarQube analysis
    \begin{itemize}
        \item The moderator presented the results of the SonarQube analysis to the participants.                                \item Discussion on the results of the SonarQube analysis
        \end{itemize}
    \item Wrap-Up                                                                                               
\end{itemize}


\subsubsection{Research Instrument}
The Focus group was based on four tasks. Each task was driven by a Question:
\begin{itemize}
    \item [Q1] What benefits and issues have you perceived after the migration to microservices?  
    \item [Q2] Is the quality of the code changed after the migration to microservices?  

    \item [Q3] Has the TD decreased after the migration to microservices? 

    \item [Q4] Have you postponed any technical activities before or during the migration to microservices? 

\end{itemize}

\end{appendices}
\end{document}